\documentclass[conference]{IEEEtran}
\IEEEoverridecommandlockouts
\usepackage{cite}
\usepackage{amsmath,amssymb,amsfonts}
\usepackage{algorithmic}
\usepackage{graphicx}
\usepackage{textcomp}
\usepackage{xcolor}
\usepackage{hyperref}
\def\BibTeX{{\rm B\kern-.05em{\sc i\kern-.025em b}\kern-.08em
    T\kern-.1667em\lower.7ex\hbox{E}\kern-.125emX}}
\begin{document}

\title{Computational anatomy atlas using multilayer perceptron with Lipschitz regularization\\
{}
\thanks{This work has been supported by the grants the Russian Science Foundation, RSF 22-21-00930.}
}

\author{\IEEEauthorblockN{Konstantin Ushenin$^{1, 2}$}
\IEEEauthorblockA{\textit{$^{1}$Institute of Immunology and Physiology} \\
\textit{$^{2}$Ural Federal University}\\
Ekaterinburg, Russia \\
0000-0003-0575-3506}

\and

\IEEEauthorblockN{Maksim Dzhigil$^{1}$}
\IEEEauthorblockA{\textit{$^{1}$Ural Federal University}\\
Ekaterinburg, Russia \\
maxdzhigil@gmail.com}

\and

\IEEEauthorblockN{Vladislav Dordiuk$^{1, 2}$}
\IEEEauthorblockA{\textit{$^{1}$Institute of Immunology and Physiology} \\
\textit{$^{2}$Ural Federal University}\\
Ekaterinburg, Russia \\
vladislav0860@gmail.com}
}

\maketitle

\begin{abstract}
A computational anatomy atlas is a set of internal organ geometries. It is based on data of real patients and complemented with virtual cases by using a some numerical approach. Atlases are in demand in computational physiology, especially in cardiological and neurophysiological applications. Usually, atlas generation uses explicit object representation, such as voxel models or surface meshes. In this paper, we propose a method of atlas generation using an implicit representation of 3D objects. Our approach has two key stages. The first stage converts voxel models of segmented organs to implicit form using the usual multilayer perceptron. This stage smooths the model and reduces memory consumption. The second stage uses a multilayer perceptron with Lipschitz regularization. This neural network provides a smooth transition between implicitly defined 3D geometries. Our work shows examples of models of the left and right human ventricles. All code and data for this work are open.
\end{abstract}


\section{Introduction}

A computational anatomy atlas is a collection of geometry models. These models are organs or organ parts. The atlas is usually built on data of some patients and generates additional shapes using a numerical method. Most research is focused on cardiac atlases \cite{ordas2007computational} or brain atlases \cite{ma2010bayesian}. In computational cardiology, the atlases extend the number of simulation cases. In the neurological sciences, atlases usually help with measurements of variability for complex 3D structures. The atlases also can produce shapes of personalized implants for surgery replacements of the vertebrae and joints.

The most general approach to the computational atlas is a diffeomorphism that map some vector $\mathbf{x} \in \mathbb{R}^{m}$ to closed manifold $\phi(\mathbf{x})$. Where $\mathbf{x}$ is a simplified representation of objects. Most approaches define $\phi$ in such a way that $\mathbf{x}$ becomes an analog of principal components and $\phi(\mathbf{x})$ is data in high-dimensional space. Thus, the embedding $\phi$ allows a researcher to measure the distance between shapes and generate new shapes using some operations with $\mathbf{x}$ instead of working with original geometries \cite{tward2016parametric}. Another approach is to reduce of the atlas generation problem to the shape interpolation problem. The last one is being studied in computer graphic applications \cite{solomon2015convolutional}.

Usually, 3D objects are presented as voxel models, point clouds, or surface meshes. These approaches are known as explicit representation. Implicit representation presents an object in form of function $F(x,y,z)=s$. The object's surface is an isosurface of zero level $F(x,y,z)=s=0$. This definition is more useful when $s$ is a signed distance function (SDF) of the object. In this case, $s$ shows a distance to the object. $s>0$ if $(x,y,z)$ is located outside the object, and $s<0$ if $(x,y,z)$ is located inside the object. SDF may be approximated with its value on a uniform grid. This is known as a signed distance field. To avoid ambiguous notation for function and field, we denote a 3D array of distances to the object surface as a signed distance tensor (SDT) instead of the signed distance field. The other approach, is to approximate of SDF with some method of function approximation, for example with multilayer perceptron (MLP).

Recent advances in implicit representation of 3D objects is a multilayer perceptron with Lipschitz regularization (LipMLP) \cite{liu2022learning}. This approach uses a special loss to force MLP to satisfy the Lipschitz condition: $\| \operatorname{MLP}(\mathbf{x}_1) - \operatorname{MLP}(\mathbf{x}_2) \|_2 < c \| \mathbf{x}_1 - \mathbf{x}_2 \|_2$, where $c$ is a positive constant. If this neural network is trained to the implicit representation of 3D objects, then it can produce a smooth transition between objects on inference.

In this paper, we propose an approach to building a computational anatomy atlas. Here the atlas generation problem is reduced to the shape interpolation problem. The first stage of this work converts the voxel model to a discrete signed distance field and then converts this tensor to a signed distance function (SDF) using the usual MLP. The second stage trains the MLP with Lipschitz regularization (LipMLP) to provide a smooth transition between SDFs.

\section{Methods}

For this study, we use a \textbf{training dataset} of Automated Cardiac Diagnosis Challenge (ACDC) \cite{bernard2018deep}. This dataset includes computed tomography data of 100 patients and ground truth segmentation of the right ventricle, left ventricle, and ventricular myocardium. We choose the segmented region of the ventricular myocardium from models: 'patient 9', 'patient 29', and 'patient 89'. The cardiac anatomy of 'patient 9' is a normal case, 'patient 29' is a myocardial infarction case, and 'patient 89' is an abnormal right ventricle case.

The original data are a 3D voxel models of the left and right myocardium. Spatial resolutions are not equal along the axis. To fix this, we apply bicubic interpolation along the Z-axis and reconstruct the real shape of the object. Smoothing methods are not applied. After interpolation, we convert voxel models to the 3D array of the signed distance field. All these transformations and methods are available in the \textit{scipy} package for Python. To avoid ambiguous notation for function and field, we denote a 3D array of distances to the object surface as a signed distance tensor (SDT) instead of the signed distance field or signed distance function field. 

\textbf{In the first part of study}, we aim to transform SDT to signed distance function (SDF) in form $F(x,y,z)=s$. For sake of simplicity, assume that MLP includes $d$ hidden layers with $w$ neurons in each. The first layer has an input size equal to three. The last layer has an output size equal to one. Such neural networks have the form of:

\begin{align}
     \mathbf{x}^{(1)}_{w \times 1} &= \sigma ((x,y,z) \mathbf{W}^{(1)}_{w \times 3}  + \mathbf{b}^{(1)}_{w \times 1})\\
     \mathbf{x}^{(i+1)}_{w \times 1} &= \sigma (\mathbf{x}^{(i)}_{w \times 1} \mathbf{W}^{(i)}_{w \times w}  + \mathbf{b}^{(i)}_{w \times 1})\\
     s &= \sigma (\mathbf{x}^{(l)}_{w \times 1} \mathbf{W}^{(l+1)}_{1 \times w}  + \mathbf{b}^{(l+1)}_{1 \times 1}),
\end{align}
where $\sigma=\operatorname{relu}(\cdot)$ is an activation function, $\mathbf{W}^{(i)}, i \in [1,l]$ is a weight matrix, $\mathbf{b}^{(i)}, i \in [1,l]$ is a bias vector, and $\mathbf{x}^{(i)}_{w \times 1}$ is an output of an $i$-th layer.

The MLP are trained with mean square error loss: $L = (1/N)\| \operatorname{MLP}(x,y,z) - \operatorname{SDT}([x],[y],[z]) \|_{2}$, where $[\cdot]$ is an operation of coordinate conversion to the index of the SDF array. Models are trained by the ADAM algorithm with a 0.006 learning rate. A coordinate set of points ($\{(x,y,z)\}$) is defined as a uniform grid inside a minimal bounding box that includes all objects. We use a uniform grid with a size 128x128x128, which corresponds to 2,097,152 points.

Note that SDT to SDF conversion is also a transfer of the explicit object representation to the implicit object representation. The result section of this paper includes some theoretical considerations about the algorithmic complexity of objects with SDT and MLP. 

In this paper, we evaluate various depths and widths of the MLP, aiming to find a small but efficient neural network for the implicit object representation. MLP is trained on SDT data of 'patient 9'. Inference of MLP provides SDF values with high spatial resolution.

\textbf{In the second part of the study}, we use MLP to approximate function $G(x,y,z,\mathbf{p})=s$. Vector $\mathbf{p} = (p_1, p_2,..,p_m)$ is one-hot encoding of model number in the set of cases. Training of LipMLP requires some set of $(x,y,z)$ points and a $\mathbf{p}$ vector. During the training, the $\mathbf{p}$ vector includes 1 value on only one position.  

The loss for LipMLP is $L = (1/N)\| \operatorname{MLP}(x,y,z,\mathbf{p}) - \operatorname{SDT}_n([x],[y],[z]) \|_{2} + \lambda \prod_{i=1}^{l} \|W^{(i)}\|_{\infty}$, where $n$ is an index of model that encoded in $\mathbf{p}$. We set $\lambda = 1e-6$. Implementation of LipMLP is adopted from the source code of \cite{liu2022learning}.

If the neural network is trained on such data, then it can produce the computational anatomy atlas mixing the initial geometries in various proportions. Set of unit-sized $\mathbf{p}$ vectors define $m$-simplex. Any vector inside the simplex defines a new model for the computational anatomy atlas. Vector $\mathbf{r} = (r_1,r_2,..,r_m)$ is located in the simplex if and only if sum of $r_i$ equals to one ($\sum_{i=1}^{m} r_i = 1$).

Generalized barycentric coordinates \cite{floater2015generalized} can define a uniform grid in such a simplex. The uniform grid in 1-simplex is a usual linear interpolation between two original shapes. The uniform grid in 2-simplex is equivalent to the ternary plot. High-order simplex can be used, but such visualization is not so straightforward, because models cannot be properly arranged on 2D images. However, any user can mix original 3D objects in any proportions while the $\sum_{i=1}^{m} p_i = 1$ condition is satisfied.

We use the marching cube algorithm with a 128x128x128 grid to visualize all implicit representations in this study.

\section{Results}

\subsection{Theoretical consideration}

SDF ($F(x,y,z)=s$) is an abstract mathematical object. Working with it requires some data structure. SDT is a zero-order approximation of SDF around the 3D object. MLP/LipMLP is the neural network that approximates SDF on some set of points. From this point of view, SDT and MLP are data structures that present SDF inside a computer.

Access to an element of tensor requires $O(1)$ time. Reconstruction of SDF value with equal spatial resolution in all directions requires $O(k^3)$ memory, where $k$ is a number of elements in a uniform grid that is defined inside a minimal bounding box around the object. MLP with depth $d$ and width $w$ includes $(d-1)(w+1)^2 + 4w$ trainable parameters and performs $(2w^2+1)(d-1) + 8w + 2$ operations on the forward pass. Thus, the implicit representation of an object with MLP requires $O(d w^2)$ memory. The cost of SDF computation is $O(dw^2)$. However, all modern computational accelerators can use big mini-batch ($w << N,\ d << N$) and data transfer channels with bandwidths greater than $N$. For such conditions, the average time to access one SDF value is empirically close to $O(1)$.

SDF approximation with SDT, $k=64$, and \textit{float32} precision requires 1,048,576 byte. SDF approximation with MLP $w=64,\ d=8$, and \textit{float32} precision requires 238,648 bytes. MLP approximately requires 23\% of memory for the presentation of the same object. The time to obtain one SDF value depends on the implementation, but it is approximately equal to $O(1)$ for systems where computation time is cheaper than data transfer. Note that all theoretical considerations are correct for LipMLP because this neural network works like a usual MLP on the inference.

\subsection{Implicit representation of one object}

\begin{figure}[ht]
\includegraphics[width=0.5\textwidth]{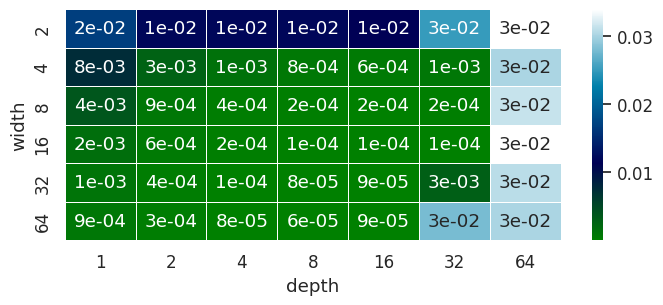}
\centering
\caption{Minimal loss is reached during 10,000 iterations of training. Rows and columns show depth and width of the MLP. 8 layers with 64 neurons provides the best results.}
\label{fig:width-heigth}
\end{figure}

\begin{figure}[ht]
\includegraphics[width=0.5\textwidth]{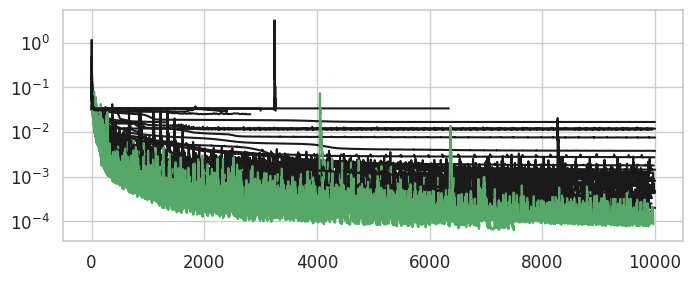}
\centering
\caption{Training history of all MLPs that provide implicit representation of the object. Green lines show the training history that reaches minimal loss below the 10e-5 threshold. Some training finished  with early stopping rules because of high loss or gradient explosion.}
\label{fig:history}
\end{figure}

Figure \ref{fig:width-heigth} shows the minimal loss that is reached by the MLPs. An MLP with 64 neurons in each of the 8 layers provides the best approximation of SDT. Figure \ref{fig:history} show the training history for each MLP. There are many MLPs that show good performance in the approximation of SDT. Virtually, any MLP with 2-16 hidden layers and 4-64 neurons in each layer is suitable for the implicit representation. MLPs with two neurons per layer are too narrow and cannot represent the object properly. This can be observed in the first row in Figure \ref{fig:width-heigth}. Neural networks that are too deep have an issue with the descent gradient. This can be seen from the last column in Figure \ref{fig:width-heigth}. According to Figure \ref{fig:history}, shallow and narrow models cannot reduce loss below some lower boundary. Gradient explosion is also presented in such models. Models that outperform the 10e-5 threshold in 10,000 iterations can be trained further. However, we stop the training because better performance is not necessary for the goals of this study.

\begin{figure}[ht]
\includegraphics[width=0.5\textwidth]{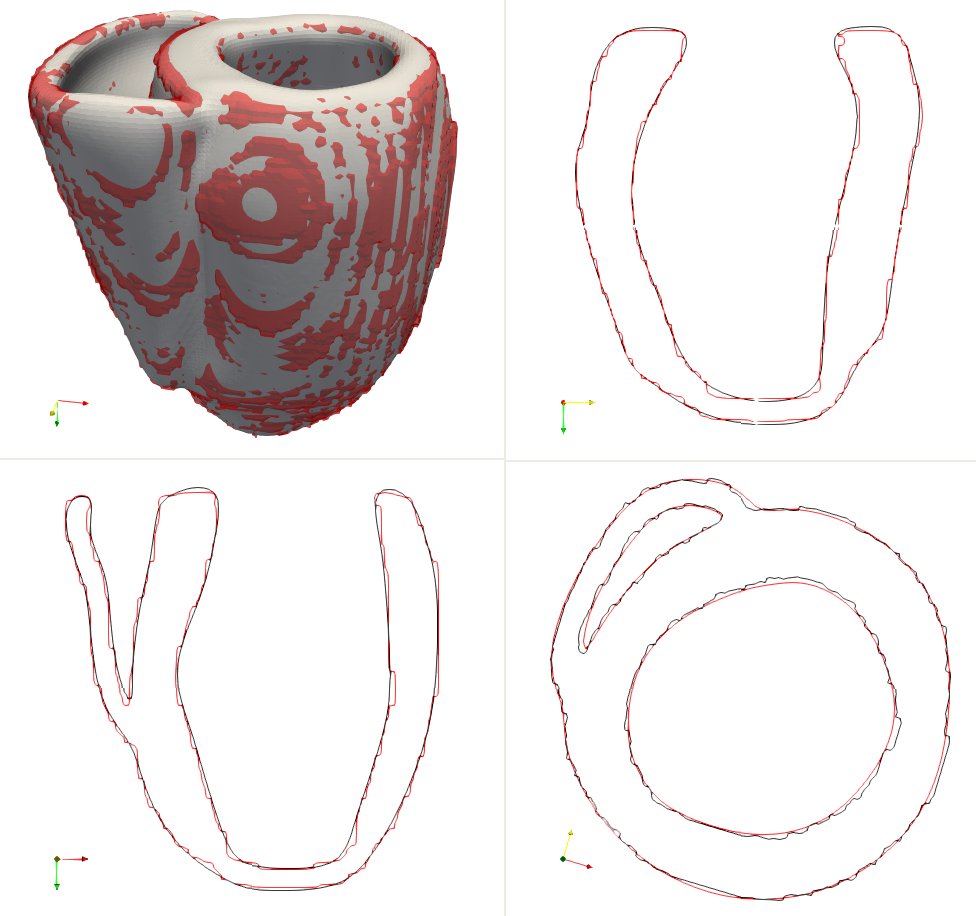}
\centering
\caption{Two representations of the object: SDT (red) and SDF (black or white). Note the smoothness effect of the implicit representation with MLP.}
\label{fig:smooth}
\end{figure}

Figure \ref{fig:smooth} shows the object reconstructed from SDT and SDF with the cube marching algorithm. We obtain these results for the best MLP scheme (8 hidden layers, 64 neurons per layer). The strong smoothing effect of the conversion is clearly visible. The implicit representation reproduces all important physiological features, such as wall thickness, places of left and right ventricle fusion.

\subsection{Computation anatomy atlas}

\begin{figure}
\includegraphics[width=0.5\textwidth]{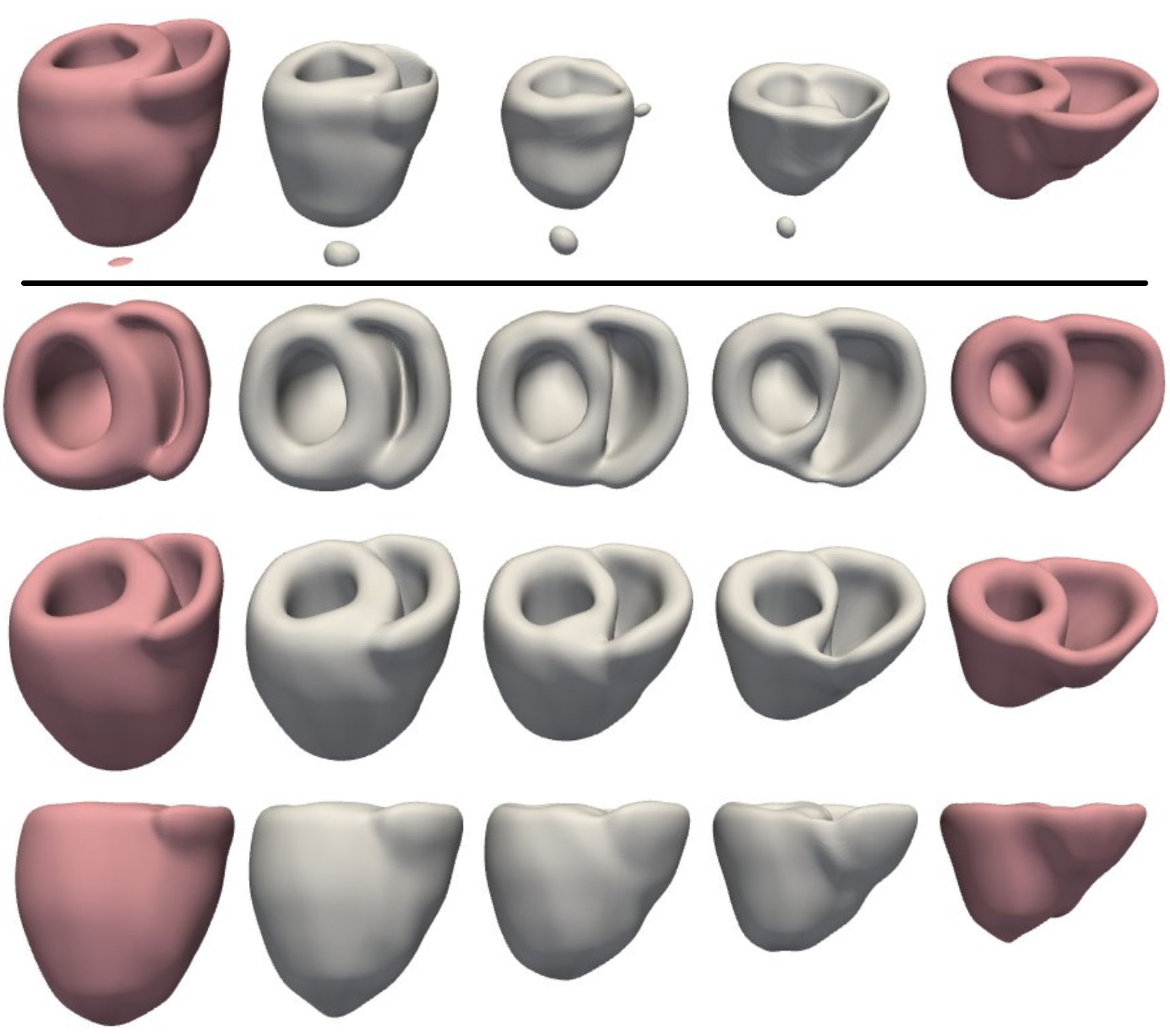}
\centering
\caption{Shape interpolation between two objects. Results for usual MLP are above the black line. Results for LipMLP are below the black line.}
\label{fig:comparison}
\end{figure}

\begin{figure}
\includegraphics[width=0.5\textwidth]{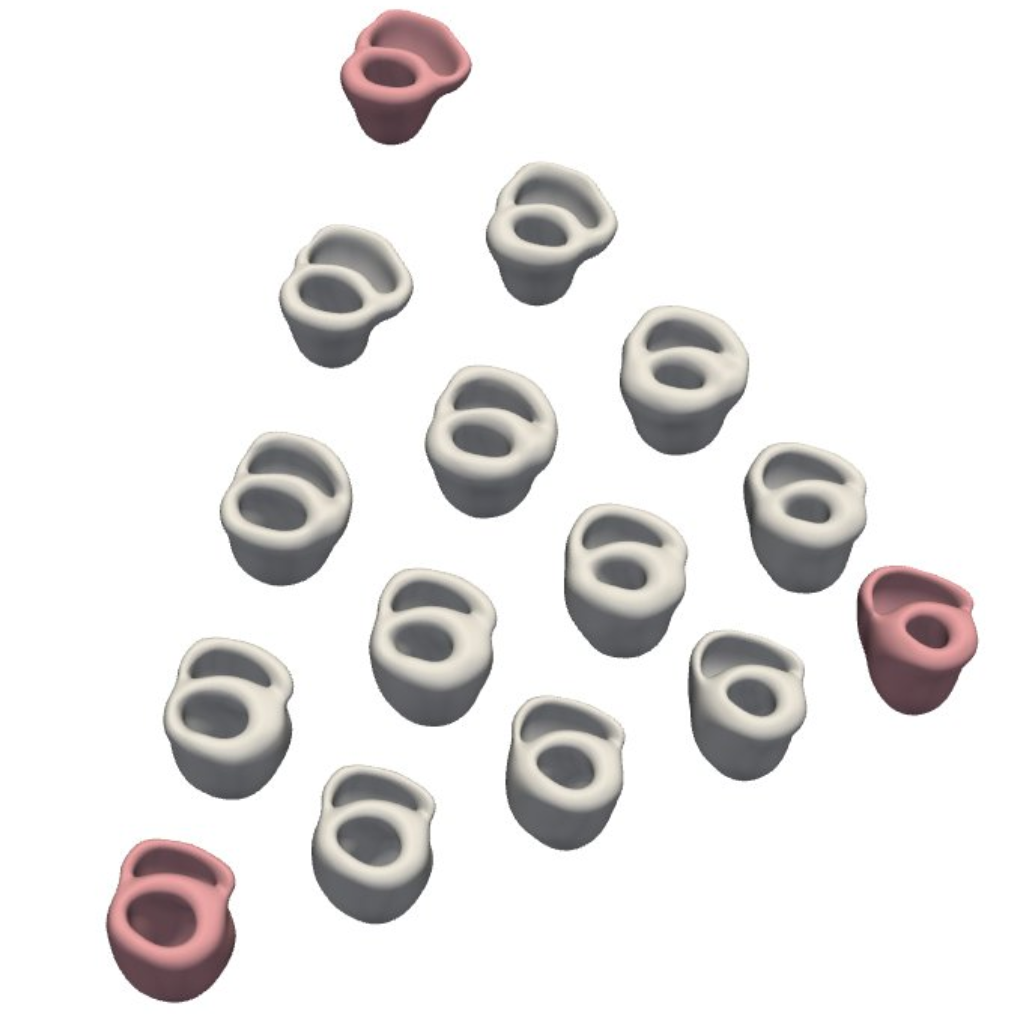}
\centering
\caption{Shape interpolation between three objects with LipMLP.}
\label{fig:comparison_1}
\end{figure}

To build the computational anatomy atlas, we uses interpolation between two cases ('patient 9', 'patient 89') and three cases ('patient 9', 'patient 29', 'patient 89'). Figure \ref{fig:comparison} shows the shape interpolation between ventricles of 'patient 9' and 'patient 89'. The upper row shows the interpolation with MLP. Results are poor and include some blobs outside the expected objects. Other rows show the interpolation with LipMLP. The transition between the two shapes is rather pretty smooth, as can be seen. Figure \ref{fig:comparison_1} shows an anatomical atlas based on the three models.

\section{Discussion and Conclusion}
 
In this work, we propose a new approach to building the computational anatomy atlas. Our approach employs an implicit neural representation of 3D objects from \cite{liu2022learning}. The first step is to convert the original voxel model to SDF. The conversion provides smoothing for the rough voxel model and reduces memory consumption for further operations. The second step is to train LipMLP on the set of objects and to infer LipMLP with parameters that mix the original geometries in some proportions. Thus, the problem of building an anatomical atlas is reduced to the problem of shape interpolation, which is solved with neural networks.

\section{Source Code}

Examples are available as part of the Neural Network for Digital Twin (NNDT) package:  \url{https://github.com/nndt-team/nndt/tree/main/examples}. NNDT is our experimental open source library that implements some methods for computational anatomy and physiology with the neural network. Note that we are working on the library and some APIs will be changed in further releases.

\bibliographystyle{IEEEtran}
\bibliography{references}

\end{document}